\documentclass{article}

\usepackage{arxiv}

\usepackage[utf8]{inputenc} 
\usepackage[T1]{fontenc}    
\usepackage{hyperref}       
\usepackage{url}            
\usepackage{booktabs}       
\usepackage{amsfonts}       
\usepackage{nicefrac}       
\usepackage{microtype}      
\usepackage{lipsum}
\usepackage{graphicx}

\title{When Physical Unclonable Function Meets Biometrics}

\author{
  Kavya~Dayananda \\
  Department of Computer Engineering\\
  San Jose State University\\
  San Jose, CA 95119 \\
  \texttt{kavya.dayananda@sjsu.edu} \\
    \And
  Nima~Karimian \\
  Department of Computer Engineering\\
  San Jose State University\\
  San Jose, CA 95119 \\
  \texttt{nima.karimian@sjsu.edu} \\
}

\begin{document}
\maketitle

\begin{abstract}
As the Covid-19 pandemic grips the world, healthcare systems are being reshaped, where the e-health concepts become more likely to be accepted. Wearable devices often carry sensitive information from users which are exposed to security and privacy risks. Moreover, users have always had the concern of being counterfeited between the fabrication process and vendors' storage. Hence, not only securing personal data is becoming a crucial obligation, but also device verification is another challenge. To address biometrics authentication and physically unclonable functions (PUFs) need to be put in place to mitigate the security and privacy of the users. Among biometrics modalities, Electrocardiogram (ECG) based biometric has become popular as it can authenticate patients and monitor the patient's vital signs. However, researchers have recently started to study the vulnerabilities of the ECG biometric systems and tried to address the issues of spoofing. Moreover, most of the wearable is enabled with CPU and memories. Thus, volatile memory-based (NVM) PUF can be easily placed in the device to avoid counterfeit. However, many research challenged the unclonability characteristics of PUFs. Thus, a careful study on these attacks should be sufficient to address the need. In this paper, our aim is to provide a  comprehensive study on the state-of-the-art developments papers based on biometrics enabled hardware security. 
\end{abstract}


\section{Introduction}
As the Covid-19 pandemic grips the world, Telehealth, and Telemedicine play crucial roles to provide a better quality of care and patient monitoring as they do not rely on in-person services. According to CDC~\cite{CDC}, Telemedicine minimizing the spread of SARS-CoV-2 while it provides necessary care to patients. Although there are many benefits of healthcare application, there has always been a threat of hackers getting access to the sensitive information of the patients that can make misuse that information leading to problems~\cite{tehranipoor2018low,limaye2018hermit}. Telemedicine and its intersection with wearable devices accelerate patient insights. Therefore, the increased usage of biometric technologies raises security and patient privacy risks, especially in unprotected and hostile environments. In addition, cybercriminals around the world work continuously to discover new techniques to invade the systems. There is a need for patient authentication such that it can protect the user's privacy and their sensitive personal data. To address this issue, biometric technology such as electrocardiogram (ECG) and photoplethysmogram (PPG) are the best candidates~\cite{karimian2019unlock}. Apart from its features for extracting health data insights and diagnosis through healthcare monitoring~\cite{yin19,cordeiro2020ecg}, ECG enables user authentication based on physiological signals\cite{ingale2020ecg,odinaka2012ecg,labati2019deep}. Moreover, consumer electronics such as wearable devices have always had the concern of being counterfeited between the fabrication process and vendors' storage~\cite{negka2019employing}. Hence, it is necessary for the hospital and user to verify the devices. To address this concern,  some of the existing components in the wearable devices such as memory can make a security layer on top of the system gathering signals from a patient. Physical Unclonable Functions (PUFs) are one of the hardware solutions commonly used to authenticate with any given device~\cite{mohanty2020pufchain,yanambaka2019pmsec,tehranipoor2018low}. In the continuation of this article, we will discuss current challenges of biometric, PUFs, and future possibilities in combining  two advances in physical unclonable functions (PUFs) and biometrics systems. 

\section{Background on ECG Biometric}
The work of Odinaka et al.~\cite{odinaka2010ecg} presents a unique short-time frequency method for ECG biometric. A large dataset with a sample size of 269 was obtained from the general population.
These single lead ECG samples were collected over a time period of seven months at different
days and times. Also, care was taken to choose subjects from different geographical regions and
health statuses. Additionally, the samples were collected in a manner that presented least
discomfort to the subjects by placing the electrodes bilaterally on the lower ribcage. After
collecting the data samples, filtering and preprocessing techniques were applied on them to
reduce the effect of noise on the input sample. First the signals were digitally notch filtered to
remove the interference from the power line. After filtering, the R-wave peaks of the signal are
detected and the signal is segmented and aligned with respect to the obtained R-wave peak. The
other fiducial points $P, Q, S, T$ are neither detected nor used in this method. Finally each of the ECG signals is normalized. Normalization is done by subtracting the sample mean of the pulse,
and dividing by the sample standard deviation. Finally for selecting the features, the authors
propose a novel method for feature selection based on the distinguishability and stability of
features. From each ECG pulse signal, a spectrogram value is calculated .It is computed by
considering the logarithm of the square of the magnitude of the short-time Fourier transform of a
normalized ECG heart pulse. This method achieved an accuracy of upto 93.5\%. Furthermore, the presented an 5.58\% error and 76.9\% accuracy during train-test phases for heartbeats from
different days. For data from a single day, the train-test error rate achieved was 0.37\% and 99\%
accuracy for decisions based on a single heartbeat.

Li et al.~\cite{li2010robust} propose an alternative approach to perform analysis of the signals in temporal and spatial domains. They extract ECG features from time slices that make up the signal known as spectrum. The features extracted are then used with the temporal information to derive a heterogenous model to determine the heart rate and fiducial characteristics. They achieve up to
99\% on the MIT-NIH database. They show that identification of spatio-temporal characteristics
using gaussian mixture method (GMM) and support vector machine based temporal analysis
improve the accuracy to 98\% from 89\% that uses traditional techniques. On the other hand , Labati et al.~\cite{labati2019deep} use convolutional neural networks to predict ECG signal characteristics. They use ECG signals from the ECG E-HOL-03-0202-003 database to derive the test and train data for the neural networks. They preprocess the ECG data from the database and then use it for feature extraction using the neural network. The preprocessing consists of filtering the noise and deriving a subset of the input sample called feature vector that is used for all subsequent processing. They have a 12-layer convolutional neural network to identify the features from the feature vector for the given test and train samples. This method achieved an accuracy of up to 97\% and an error rate of 2.90\%.

\section{Background on Biometric Key Generation}
\begin{figure}
    \centering
    \includegraphics[width=0.55\linewidth]{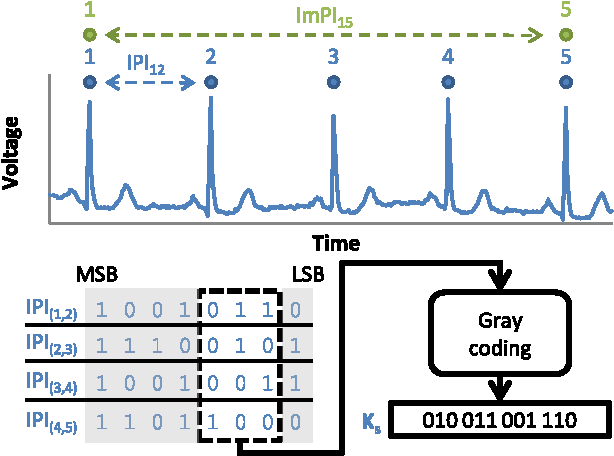}
    \caption{ECG Key generation using the interpulse intervals (IPIs)~\cite{seepers2015enhancing}.}
    \label{fig:IPI}
\end{figure}

There have been numerous proposals for generating keys from biometrics. The key requirements for biometic key is to meet the requirement such as randomness and uniqueness~\cite{ballard2008practical,karimian2016highly}. Generally, biometric features will be transformed into a binary string of various lengths to represent biometric keys. Biometric keys generated should be uniformly random in order to remain resistant
to attack. There have been many research on biometric key generation developed such fingerprint~\cite{ratha2007generating,jin2016biometric}, face~\cite{goh2003computation}, hand geometry~\cite{beenau2004method}, and iris~\cite{lee2007biometric}. Not only the aforementioned biometric keys are vulnerable to spoofing attacks, but also implementing biometric key based on iris, face is not feasible in Telemedicine. Thus, ECG based biometric key generation has been introduced. ECG based bio-key not only offers liveness and are difficult to spoof, but also it is more feasible to be implemented in Telemedicine systems as its extracting health data insights and diagnosis through healthcare monitoring. Early works proposed interpulse intervals (IPIs) to use electrocardiograms (ECGs) as an entropy source for generating random binary sequences~\cite{rostami2013heart,pirbhulal2018heartbeats,moosavi2017cryptographic,zhang2011analysis,pirbhulal2018heartbeats}. The IPI is defined as the time interval between two successive R peaks.  First steps is to clean the ECG signal from noise sources. To eliminate the noise, the ECG signal is passed through a Butterworth pass-band filter. Second,   using Pan–Tompkins algorithm, R peak will be identified. Second, the time
difference between two consecutive R-peaks are computed to generate IPIs. Each IPI will be converted into values between 0 and 1. The 128 key bits will be generated based on LSB of each IPIs. The drawback of this technique is that it requires a long period of IPIs~\cite{altop2017deriving,rostami2013heart}. The example of IPI key generation is demonstrated in the Fig~\ref{fig:IPI}. As can be seen in this figure, due to low entropy of the most-significant bits (MSB), only least significant bit (LSB) are considered and MSB bits are discarded.

While biometrics have their advantages, keys generated from biometrics may suffer from environment noise. These limitations could introduce binary errors during key generation and make it impossible for even the system owner to access/use the system. Thus, IPI techniques heavily rely on error correction. In order to overcome these detriments, a statistical approach to mitigate or eliminate the intra-subject variance while preserving privacy and generating long keys have been developed~\cite{karimian2016highly,karimian2017noise}. The statistical approach begins with an initialization phase. ECG Features are extracted from ECG signals from a large population of users. Statistics for each feature are used to determine thresholds and boundaries to quantize the feature into one or more key bits. The number of bits depends on the parameters and statistics, and can vary from feature to feature. Compared to IPI methods, statistics approach use single heartbeat to generate 128 key bits. The proposed system containing biometric pre-processing, feature extraction, etc. and reconfigurable logic to act as the permutation block. The reconfigurable logic forms is programmed according to the operators biometric. They used permutation chip to insert a binary key bits and lock the system. If the correct biometric which belongs to the authorized owner is asserted, the system will unlock allowing the inputs pass to the correct outputs of the permutation chip. Otherwise, the system will not work.

\begin{figure}
    \centering
    \includegraphics[width=0.6\linewidth]{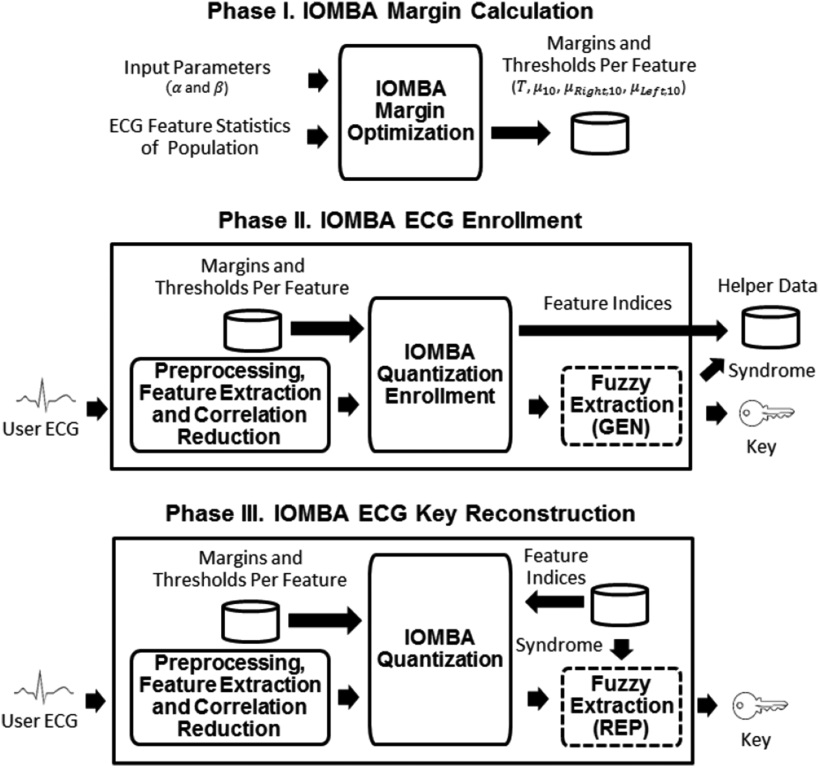}
    \caption{ECG Key generation using the statistical approach~\cite{karimian2016highly}.}
    \label{fig:IOMBA}
\end{figure}

\section{Background on PUFs}
PUFs take advantage of the unique physical characteristics of an integrated circuit (IC) that result from manufacturing and process variation in nanoscales, giving each its own fingerprint. A PUF relies on inherent entropy as well as stability/reliability to provide uniquely secure, yet consistent operation based on a unique set of challenge (input) and response (output) pairs. PUF's uniqueness involves determining whether or not it provides a different enough signature for its given IC to clearly differentiate it from other ICs of the same kind. Computing the Hamming distance (HD) between a pair of PUF identifiers (or the number of bit positions that differ in value) is one way this can be done. Reliability refers to how well a PUF is able to provide a consistent response to the same challenge. This is crucial since a PUF should not be easily worn out or affected by environmental factors.  There are many forms of existing PUFs, including delay-based PUFs such as arbiter PUFs~\cite{hospodar2012machine,avvaru2016estimating}, ring-oscillator PUFs~\cite{merli2010improving,yan2017phase,maiti2010large,chen2011bistable}; memory-based PUFs such as DRAM PUFs~\cite{tehranipoor2015dram,xiong2016run,tehranipoor2016dram,karimian2019generate} and SRAM PUFs~\cite{holcomb2008power,schrijen2012comparative,garg2014design}, etc. ``Weak and Strong PUFs" are the two subtypes of PUFs. Weaks PUFs are used for storing secret keys to non-volatile memories since they show some internal, unclonable physical disorder, and they are involved in some form of challenge-response mechanism which should be access-restricted. It is considered that even by having the PUF-carrying hardware, the adversaries cannot access the Weak PUF’s responses. Formerly, weak PUFs were toward special purpose circuits, now they are based on intrinsic PUFs built from CMOS parts which are more affordable. Strong PUFs have numerous CPRs which each time provide a new CPR for the procedure of authentication. They exhibit a more complex challenge-response pair. They are much harder to be predicted since if an adversary knows a large subset of CRPs, it is almost impossible for them to predict the other unknown CRPs. Strong PUFs have public access for challenge-response mechanisms that allow everyone with physical possession of the PUF  to apply challenges to the Strong PUF which leads to some downsides such as the need for many CRPs to remain secure~\cite{overviewmemory17}.

\begin{figure}
    \centering
    \includegraphics[width=0.7\linewidth]{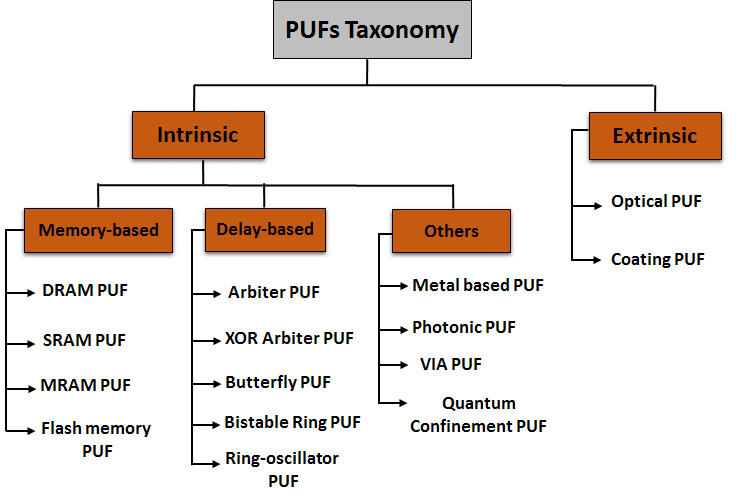}
    \caption{PUFs Taxonomy~\cite{tehranipoor2017design}.}
    \label{fig:PUFtax}
\end{figure}

\section{Combining Hardware Security and Biometric Systems}
\begin{figure}
    \centering
    \includegraphics[width=0.7\linewidth]{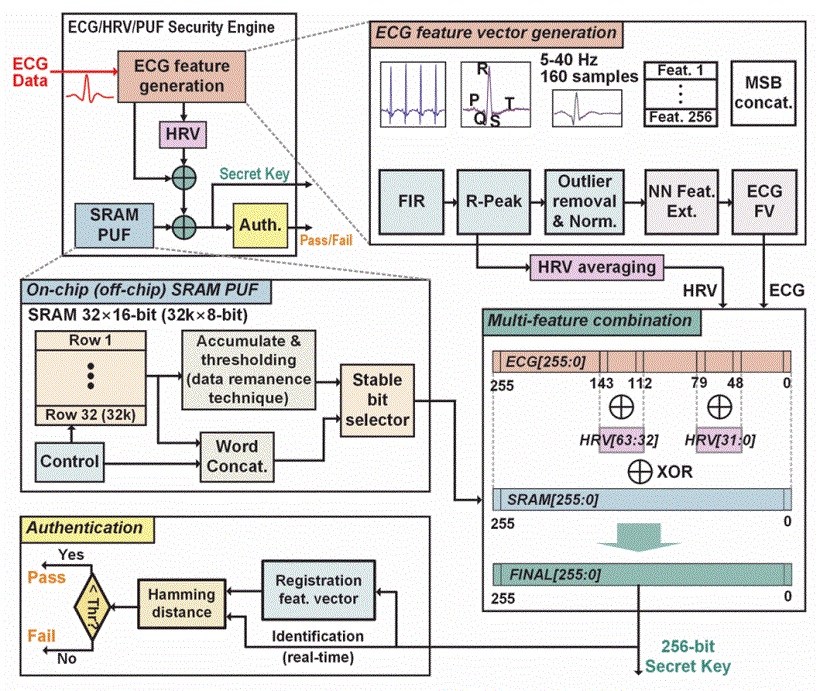}
    \caption{Combining hardware security with biometric system. Features based ECG, HRV and SRAM PUF are extracted and combined to form a 256-bit secret key~\cite{cherupally2020smart}.}
    \label{fig:PUFECG}
\end{figure}

Since unique signature of the wearable device using PUF do not vary for different users and unique signature of a user using biokeys do not vary across multiple devices, Cherupally et al.~\cite{cherupally2020smart} combined ECG biokey, heart rate variability (HRV), and SRAM-based physical unclonable function (PUF) to perform real-time authentication and generate unique/random signatures. The purpose of the proposed work was to  enhance randomness. As can be seen in Fig\ref{fig:PUFECG}, 256 key bits from individual's ECG and HRV are combined with  256 key bits of each device's PUF using bitwise XOR to generate a 256-bit secret key. Then this key will be used as one-time key for authentication. This methods will reduce attack scenario  to know both the PUF as well as the
user’s ECG/HRV features to uncover the root of trust~\cite{cherupally2020smart}.

Guo et al.~\cite{guo2016hardware} combined ECG biokey and physical obfuscation/locking to protect electronic system by physical obfuscation/locking. The purpose of the proposed work is avoid fault injection, and reverse engineering. They have considered two approaches for users enrollment. In the first approach authorized user could present his/her biometric to the system.Alternatively, a database containing biometric keys from authorized user can be used to configure the system. The second approach has the advantage that the user need not be present during enrollment. To that, the system could be enrolled in a secure location and then sent to the user. In that case, if the device is captured by unauthorized user, it could not be
used until it reaches its enrolled user.

Karimian et al.~\cite{karimian2018secure}, combined ECG key bits, physical unclonable functions (PUFs) and hardware obfuscation to  provides non-invertibility and non-linkability charchtestics of biometric systems. The purpose of using PUF is to avoid permanently storing the human biometric and/or cryptographic keys on the device. The PUF also provides a non-invertible transform to the biometric in order to protect the biometric template and make the biometric/obfuscation non-linkable to different devices. The hHardware obfuscation has been replace in the biometric matching systems. The users ownership is taken by presenting his/her biometric signal to the device. A preprocessing algorithm is applied on the received biometric to extract the binary bio key.  A PUF is used as a one-way transform on the quantized biometric. In order to generate the PUF challenge, the bio key is processed by a hardware hash function to the desired length. An obfuscated bitstream is produced that will exploit this obfuscation key to lock the system. During the authentication process, the user provides his/her biometric as input. The same pre-processing algorithm as the enrollment process is applied to generate the $bio-key$. Potential errors are corrected with the helper data. Next, the hash function creates the challenge. The $obs-key$ is then generated by injecting this challenge into the strong PUF. Different from the enrollment process, this obfuscation key is generated on the physical device instead of its mathematic model. A correct $obs-key$ unlocks the obfuscated bitstream and brings the device into functional (unlocked) mode. Without the correct key, the device will simply not work correctly. Arjona et al.~\cite{arjona2018puf} obfuscated the biometric data with PUFs based on dual-factor authentication protocol to avoid storage of secret keys for IoT application. They propose fingerprint recognition solution to identify the user and that protects biometric data with the PUF response at the sensor node.

\subsection{ECG based Biometric and Blockchain}
The advancement of microfabrication has led to every small device having its own way of
establishing network connectivity and access to the internet. This ecosystem of interconnected
devices is known as the Internet of Things (IoT). IoT has paved the way for a wide range of
applications in several fields. Wearable health monitoring devices like smartwatches and Fitbit
are one such application of IoT. These portable devices use various sensors to measure vital
information of the human body like the pulse rate, heart rate, and other useful personal health
metrics. So with the easy access of ECG information from these smart devices, there has been
increasing interest in using ECG data as biometrics in a variety of applications. Also, ECG
signals, similar to fingerprints, are unique to the individual and are very difficult to alter or
replicate by external means thereby making them a perfect candidate for biometric especially for
security applications. However, a major concern for these applications is the need for
transmitting these sensitive medical data and ECG signals securely across the decentralized IoT
ecosystem. Blockchain technology provides a promising solution in this regard. Blockchain
provides a decentralized reliable and secure mechanism for transmitting the data. Furthermore ,
blockchain is immutable , thus ensuring that there is no tampering of the data. 
Blockchain has been historically used by crypto currency coin base for distributed management
of assets across the world in a safe and secured manner. Such a system has been considered in
reference to medical data so that different stakeholders such as pharmacies, doctors and health
care providers can integrate to a comprehensive and unified health care system. One such effort
is provided by Vazirani et al.~\cite{vazirani2020blockchain} towards effective healthcare record management. They create a
blockchain for pertinent data ownership by associated stakeholders. This way they allow
distributed ownership of patient data without compromising on privacy. For example : the lab
testing units own the measured data , whereas the doctors will own the medicine prescription and
the pharmacist will monitor the prescription data to pack the medicines for the patient. This way,
they securely decentralize the patient’s medical history and use the block chain as a distributed
ledger to make medical history as a unifying factor for the british healthcare system. In their
study , they have shown that such a distributed ledger can support high throughput machine
learning based analysis for better healthcare and also provide secured personalised healthcare
pathways.

\section{Conclusions}
The advancement of microfabrication has led to every small device having its own way of
establishing network connectivity and access to the internet. This ecosystem of interconnected
devices is known as the Internet of Things (IoT). IoT has paved the way for a wide range of
applications in several fields. Wearable health monitoring devices like smartwatches and Fitbit
are one such application of IoT. These portable devices use various sensors to measure vital
information of the human body like the pulse rate, heart rate, and other useful personal health
metrics. So with the easy access of ECG information from these smart devices, there has been
increasing interest in using ECG data as biometrics in a variety of applications. Also, ECG
signals, similar to fingerprints, are unique to the individual and are very difficult to alter or
replicate by external means thereby making them a perfect candidate for biometric especially for
security applications. However, a major concern for these applications is the need for
transmitting these sensitive medical data and ECG signals securely across the decentralized IoT
ecosystem. Blockchain technology provides a promising solution in this regard. Blockchain provides a decentralized reliable and secure mechanism for transmitting the data. Furthermore , blockchain is immutable , thus ensuring that there is no tampering of the data. The below section presents different techniques to leverage ECG data for biometric and blockchain technology.
Moreover, Wearable devices often carry sensitive information from users which are exposed to security and privacy risks. Moreover, users have always had the concern of being counterfeited between the fabrication process and vendors' storage. Hence, not only securing personal data is becoming a crucial obligation, but also device verification is another challenge. To address biometrics authentication and physically unclonable functions (PUFs) need to be put in place to mitigate the security and privacy of the users.



\end{document}